\documentclass[10pt,conference]{IEEEtran}
\IEEEoverridecommandlockouts

\usepackage{amsmath}
\usepackage{amsfonts}
\usepackage{xcolor}
\usepackage{graphicx}
\usepackage{subcaption}
\usepackage{float}
\usepackage{multirow}
\usepackage{todonotes}
\usepackage{verbatim}
\usepackage{hyperref}

\usepackage{tikz}
\usepackage{pgfplots}
\usepgfplotslibrary{colorbrewer}
\pgfplotsset{compat=1.8}
\definecolor{clr1}{RGB}{27,158,119}
\definecolor{clr2}{RGB}{217,95,2}
\definecolor{clr3}{RGB}{117,112,179}
\definecolor{clr4}{RGB}{231,41,138}
\definecolor{clr5}{RGB}{102,166,30}
\definecolor{clr6}{RGB}{230,171,2}
\definecolor{clr7}{RGB}{166,118,29}
\pgfplotsset{
    cycle list={clr1,clr2,clr3,clr4,clr5,clr6,clr7},
}

\newtheorem{remark}{Remark}

\newtheorem{definition}{Definition}

\begin{document}

\title{Comparison of the FCFS and PS discipline in Redundancy Systems\\

\thanks{The work in this research abstract is supported by the Netherlands Organisation for Scientific Research (NWO) through Gravitation grant NETWORKS 024.002.003.}
}
\author{\IEEEauthorblockN{Youri Raaijmakers$^{*}$}
\IEEEauthorblockA{\textit{Department of Mathematics and Computer Science,}
\textit{Eindhoven University of Technology}, The Netherlands}
{* Corresponding author: y.raaijmakers@tue.nl.}
}

\maketitle

\begin{abstract}
We consider the c.o.c.\ redundancy system with $N$ parallel servers where incoming jobs are immediately replicated to $d$ servers chosen uniformly at random (without replacement). A job finishes service as soon as the first replica is completed, after which all the remaining replicas are abandoned. 
We compare the performance of the first-come first-served (FCFS) and processor-sharing (PS) discipline based on the stability condition, the tail behavior of the latency and the expected latency.  
\end{abstract}

\section{Introduction}

Nowadays, there is an abundance of cloud computing platforms that process vast numbers of jobs and consist of tens of thousands of servers~\cite{H-OPQT}. Computer system designers always make an effort to reduce the latency, as it is shown that only $400$ milliseconds of artificial delay into Google search already causes the users to perform $0.74$\% fewer searches after $4$-$6$ weeks~\cite{B-SMGS}. 

Redundancy scheduling is proposed as one of the techniques to reduce latency in applications with many servers. In redundancy scheduling each incoming job is replicated and allocated to multiple different servers. The job is completed as soon as the first replica finishes service after which all the other replicas are abandoned, also known as the cancel-on-completion (c.o.c.) variant. 

Over the last years several papers have been written on the c.o.c.\ redundancy system deriving expressions for key performance metrics, such as the stability condition and expected latency.
Closest related to this work is the expression for the expected latency derived in~\cite{GZDHBHSW-RLR} under the assumption of independent and identically distributed (i.i.d.) replicas, exponential job sizes and the FCFS discipline. In the same paper it is proved that the stability condition is given by $\rho := \frac{\lambda \mathbb{E}[X]}{N}<1$, where $\lambda$ is the arrival rate, $N$ the number of servers and $\mathbb{E}[X]$ the expected job size. 
In~\cite{AAJV-OSR}, under the same assumptions, the stability condition for the processor-sharing (PS) discipline is examined. In particular, it is shown that the FCFS and PS discipline yield the same stability condition. The results regarding the stability condition for the PS discipline were later extended to general job size distributions with possible dependence among the replicas~\cite{RBB-SRPS} (see also Section~\ref{sec: stability condition}). These results demonstrated that the stability condition for the FCFS and PS discipline is only the same in the specific case of exponential i.i.d.\ replicas. 

In this research abstract we compare the FCFS and PS discipline in c.o.c.\ redundancy systems with general job size distributions based on key performance metrics, such as the stability condition, tail behavior of the latency and the expected latency. It provides an overview of existing results as well as new insights. 

\section{Model description}

Consider the system with $N$ parallel servers, where jobs arrive as a Poisson process of rate $\lambda$. Each of the $N$ parallel servers has its own queue. We consider two service disciplines, namely FCFS and PS. When a job arrives, the dispatcher immediately assigns replicas to $d \leq N$ servers selected uniformly at random (without replacement). We allow the replica sizes $X_{1},\dots,X_{d}$ to be governed by some joint distribution $F_{\boldsymbol{X}}(x_{1},\dots,x_{d})$, where $X_{i}$, $i=1,\dots,d$, are each distributed as a generic random variable $X$, but not necessarily independent. 

Let us define $X_{\mathrm{min}} := \min \{ X_{1},\dots,X_{d} \}$ as the minimum of $d$ job sizes and $\tilde{\rho} := \frac{\lambda d \mathbb{E}[X_{\mathrm{min}}]}{N}$ as the load of the system. 

In the following sections we provide an overview of the known results for the stability condition and the tail behavior and present new (numerical) results for the expected latency. 

\section{Stability condition}
\label{sec: stability condition}
The exact expression for the stability condition with the FCFS discipline is still an open problem. However, in~\cite{RB-ASR} it is proved that no replication ($d=1$) gives a larger stability region than replication $(d > 1)$ for NBU distributions, whereas full replication $(d=N)$ gives a larger stability region than no replication for NWU distributions (for the definition of NBU/NWU distributions we refer to~\cite{RB-ASR}). 

For the PS discipline the stability condition is known and given by $\tilde{\rho} = \frac{\lambda d \mathbb{E}[X_{\mathrm{min}}]}{N}  < 1$, see~\cite{RBB-SRPS}. 
Observe that in the special case of i.i.d.\ replicas and a job size distribution that is NBU (NWU) we have that $d \mathbb{E}[X_{\mathrm{min}}]$ is increasing (decreasing) in $d$, see~\cite[Appendix A]{J-ERT}.
From this it follows that the stability region decreases (increases) in the number of replicas for job size distributions that are NBU (NWU).
Moreover, in~\cite{RBB-SRPS} it is conjectured that for i.i.d.\ replicas the stability region for the PS discipline is smaller (larger) than the stability region for the FCFS discipline for job size distributions that are NBU (NWU).

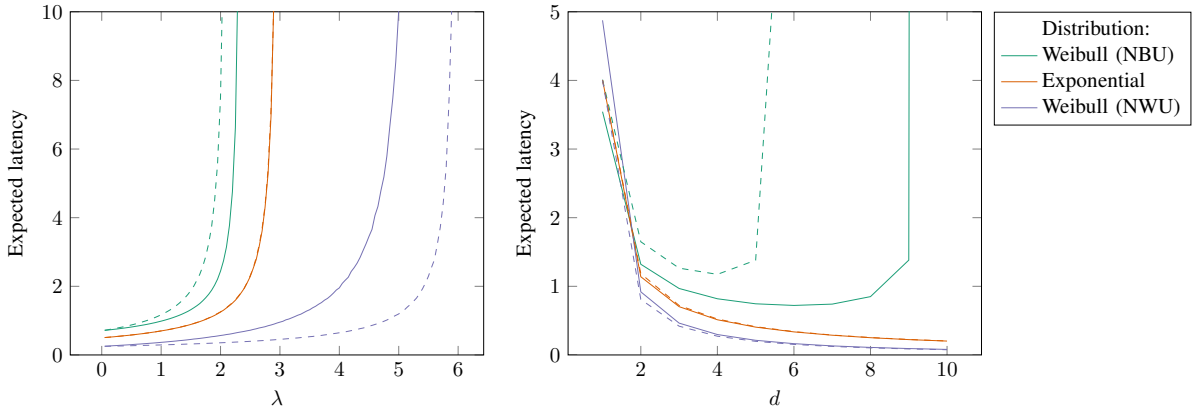
\begin{figure*}[htbp]
\centering
\resizebox{0.9\linewidth}{!}{\newcommand{\dataFigureA}{TikZFigures/ExpectedLatency_N3_FCFS.csv}
\newcommand{\dataFigureB}{TikZFigures/ExpectedLatency_N3_PS.csv}

\newcommand{\dataFigureC}{TikZFigures/ExpectedLatency_N100_FCFS.csv}
\newcommand{\dataFigureD}{TikZFigures/ExpectedLatency_N100_PS.csv}

\begin{tabular}{@{}cc@{}}
\begin{tikzpicture}
		\begin{axis}[
			xlabel=$\lambda$,
			ylabel=Expected latency,
			ymin=0,
			ymax=10,
			no markers]
		\addlegendimage{empty legend}
		\addplot+ table [x=lambda,y=WeibullNBU, col sep=comma]{\dataFigureA};
		\addplot+ table [x=lambda,y=Exp, col sep=comma]{\dataFigureA};
		\addplot+ table [x=lambda,y=WeibullNWU, col sep=comma]{\dataFigureA};
		
		\pgfplotsset{cycle list shift=-3}
		\addplot+[dashed] table [x=lambda,y=WeibullNBU, col sep=comma]{\dataFigureB};
		\addplot+[dashed] table [x=lambda,y=Exp, col sep=comma]{\dataFigureB};
		\addplot+[dashed] table [x=lambda,y=WeibullNWU, col sep=comma]{\dataFigureB};
		\end{axis}
\end{tikzpicture}
&
\begin{tikzpicture}
		\begin{axis}[
			xlabel=$d$,
			ylabel=Expected latency,
			ymin=0,
			ymax=5,
			legend pos= outer north east,
			legend cell align=left]
		\addlegendimage{empty legend}
		\addplot+ table [x=d,y=WeibullNBU_1.2, col sep=comma]{\dataFigureC};
		\addplot+ table [x=d,y=Exp, col sep=comma]{\dataFigureC};
		\addplot+ table [x=d,y=WeibullNWU_0.8, col sep=comma]{\dataFigureC};
		
		\pgfplotsset{cycle list shift=-3}
		\addplot+[dashed] table [x=d,y=WeibullNBU_1.2, col sep=comma]{\dataFigureD};
		\addplot+[dashed] table [x=d,y=Exp, col sep=comma]{\dataFigureD};
		\addplot+[dashed] table [x=d,y=WeibullNWU_0.8, col sep=comma]{\dataFigureD};
		\addlegendentry{Distribution:}
		\addlegendentry{Weibull (NBU)}
		\addlegendentry{Exponential}
		\addlegendentry{Weibull (NWU)}
		\end{axis}
\end{tikzpicture}
\end{tabular}}
\caption{Expected latency for the FCFS (solid lines) and PS (dashed lines) discipline in the scenario with $N=3$ servers and $d=2$ (left) and $N=100$ servers  and $\lambda=75$ (right) in both cases $\mathbb{E}[X]=1$, homogeneous servers and various distributions.}
\label{fig: expected latency}
\end{figure*} 

\section{Tail behavior}
Let us first introduce two classes of heavy-tailed distributions.
\begin{definition}
$X$ is $\mathcal{O}$-regularly varying, denoted by $X \in ORV$, if 
\begin{align*}
0 < \liminf_{x \rightarrow \infty} \frac{\bar{F}_{X}(\alpha x )}{\bar{F}_{X}(x)} \leq \limsup_{x \rightarrow \infty} \frac{\bar{F}_{X}(\alpha x)}{\bar{F}_{X}(x)} < \infty, ~~~ \forall \alpha \geq 1,
\end{align*}
where $\bar{F}_{X}(x) := 1-F_{X}$ is defined as the complementary cumulative distribution function. 
Furthermore, $X \in ORV(-\nu)$ if
\begin{align*}
c_{1} \alpha^{-\nu} < \liminf_{x \rightarrow \infty} \frac{\bar{F}_{X}(\alpha x )}{\bar{F}_{X}(x)} \leq \limsup_{x \rightarrow \infty} \frac{\bar{F}_{X}(\alpha x)}{\bar{F}_{X}(x)} < c_{2} \alpha^{-\nu},
\end{align*}
for all $ \alpha \geq 1$ with positive constants $c_{1}$ and $c_{2}$.
\end{definition}

\begin{definition}
$X$ is regularly varying of index $-\nu$, denoted by $X \in RV(-\nu)$, if
\begin{align*}
\bar{F}_{X}(x) = L(x) x^{-\nu}, ~~~x>0,
\end{align*}
with $L(x)$ a slowly varying function, i.e., $L(\alpha x)/L(x) \rightarrow 1$ for any $\alpha >0$ as $x$ approaches infinity.
\end{definition}
Observe that $RV \subset  ORV $, see for example~\cite[Theorem 2.1.8]{BGT-RV}.

For the FCFS discipline we have that if $X_{\mathrm{min}} \in RV(-\tilde{\nu})$ then $R \in ORV(1-\tilde{\nu})$ (see~\cite{RBB-RSHT}), whereas for the PS discipline we have that if $X_{\mathrm{min}} \in RV(-\tilde{\nu})$ then $R \in ORV(-\tilde{\nu})$ (see~\cite{RBB-STBPS}). 
These results indicate that for heavy-tailed job size distributions the PS discipline always has better tail behavior than the FCFS discipline for all dependency structures between the replicas. 


\section{Expected latency}
\label{sec: expected latency}
At present almost nothing is known about the expected latency in the case of generally distributed job sizes. Only for the (trivial) cases of no replication ($d=1$) and full replication ($d=N$) we can derive expressions for the expected latency, since in these cases the system is equivalent to an $M/G/1$ queue.

For no replication ($d=1$) and full replication ($d=N$), the expected latency for the FCFS discipline is given by
\begin{align}
\mathbb{E}[T_{\mathrm{FCFS}}] = \frac{\tilde{\rho} \mathbb{E}[X_{\mathrm{min}}^{2}]}{2(1-\tilde{\rho})\mathbb{E}[X_{\mathrm{min}}]} + \mathbb{E}[X_{\mathrm{min}}].
\label{eq: expected latency FCFS}
\end{align}

For no replication ($d=1$) and full replication ($d=N$), the expected latency for the PS discipline is given by
\begin{align}
\mathbb{E}[T_{\mathrm{PS}}] = \frac{\mathbb{E}[X_{\mathrm{min}}]}{1-\tilde{\rho}}.
\label{eq: expected latency PS}
\end{align}

\begin{remark}
When comparing the above expressions of the expected latency it can be derived that for both the FCFS and PS discipline no replication (full replication) is better if
$N \mathbb{E}[\min \{ X_{1},\dots,X_{N} \}] \geq (\leq) \mathbb{E}[X]$\footnotemark.
\end{remark}

\begin{remark}
According to Pollaczek-Khinchin we have that
\begin{align*}
\mathbb{E}[T_{\mathrm{FCFS}}] \leq (\geq) \mathbb{E}[T_{\mathrm{PS}}],
\end{align*}
if the coefficient of variation $C_{\mathrm{v}}^{2}:= \frac{\sqrt{\mathbb{E}[X^{2}]-(\mathbb{E}[X])^{2}}}{\mathbb{E}[X]} \leq (\geq) 1$\footnotemark[\value{footnote}].
\end{remark}
Moreover, a distribution that is NBU (NWU) has $C_{\mathrm{v}}^{2} \leq (\geq) 1$\footnotemark[\value{footnote}], see for example~\cite{MP-CDRT}.

\footnotetext{with equality in case of exponentially distributed job sizes}

In Figure~\ref{fig: expected latency} it can be seen that for $d=2$ we have that $\mathbb{E}[T_{\mathrm{FCFS}}] < (>) \mathbb{E}[T_{\mathrm{PS}}]$ for NBU (NWU) distributions\footnote{for exponentially distributed job sizes with $1<d<N$ the expected latency for the FCFS and PS discipline is not exactly equal}, which is in line with the results for no replication and full replication. 
For NWU distributions extensive simulation experiments suggest that full replication achieves the best performance, in terms of expected latency. 

The one-line conclusion of this research abstract, based on the stability, the tail behavior of the latency and the expected latency, is that for c.o.c.\ redundancy the FCFS discipline gives better performance for NBU distributions, whereas the PS discipline gives better performance for NWU distributions. Now, as pointed out in~\cite{H-OPQT}, the job size distribution in practice is mostly heavy-tailed with $C_{\mathrm{v}}^{2} >> 1$. We observe that, in this scenario, redundancy scheduling significantly improves the performance. In addition employing the PS discipline instead of the FCFS discipline enhances this improvement even more. 

\newpage

\bibliographystyle{plain}
\bibliography{references}

\end{document}